\documentclass[preprint,12pt,nofootinbib]{revtex4-1}
\usepackage[paper=letterpaper,margin=1.5in]{geometry}

\usepackage{graphicx}
\usepackage{latexsym}
\usepackage{amsmath, amssymb, amsfonts}
\begin{document}%

\begin{flushright}
{\normalsize
SLAC-PUB-16497\\
DESY 16-056\\
LCLS-II TN-16-08\\
March 2016}
\end{flushright}

\vspace{.4cm}


\title{Analytical Formulas for Short Bunch Wakes\\ in a Flat Dechirper \footnote[1]{Work supported in part by the U.S. Department of Energy, Office of Science, Office of Basic Energy Sciences, under Contract No. DE-AC02-76SF00515
} }

\author{Karl Bane\footnote[2]{kbane@slac.stanford.edu} and Gennady Stupakov}
\affiliation{SLAC National Accelerator Laboratory, 2575 Sand Hill Road, Menlo Park, CA 94025}
\author{Igor Zagorodnov}
\affiliation{Deutsches Electronen-Synchrotron, Notkestrasse 85, 22603 Hamburg, Germany}

\begin{abstract}
We develop analytical models of the longitudinal and transverse wakes, on and off axis for realistic structures, and then compare them with numerical calculations, and generally find good agreement. These analytical ``first order" formulas approximate the droop at the origin of the longitudinal wake and of the slope of the transverse wakes; they represent an improvement in accuracy over earlier, ``zeroth order" formulas. 
In example calculations for the RadiaBeam/LCLS dechirper using typical parameters, we find a 16\% droop in the energy chirp at the bunch tail compared to simpler calculations. With the beam moved to 200~$\mu$m from one jaw in one dechiper section, one can achieve a 3~MV transverse kick differential over a 30~$\mu$m length.

\end{abstract}

\maketitle

\subsection*{Introduction}

The corrugated, metallic beam pipe has been proposed as a ``dechirper" for linac-based X-ray FELs~\cite{Bane12}. The idea is to install this passive device in the beam line at the end of acceleration, in order to remove residual energy chirp in the beam before it enters the undulator for lasing.  Several corrugated dechirpers have been built and tested~\cite{Harrison13}-\cite{dechirperpaper}, as well as one based on dielectric-lined structures~\cite{Antipov14}. For adjustability, a dechiper section is built as two flat, corrugated plates with the beam meant to pass in between. A complete dechirper unit normally comprises two identical sections, with one orientated horizontally and the other vertically, in order to cancel the unavoidable quad wake that is excited and can lead to emittance growth. The RadiaBeam/LCLS dechirper has recently become the first one that has been tested at high energies (multi GeV) and short bunch lengths (10's of microns)~\cite{dechirperpaper}. 

The calculation of wakes of corrugated structures, assuming small corrugations and using a perturbation approach, has been performed for round structures~\cite{BaneNovo99}, \cite{BaneStupakov00}, and also for flat geometry~\cite{BaneStupakov03}.
In the case the corrugation parameters are not small compared to the aperture, as is true for dechirpers that have been built (assuming the nominal aperture), numerical methods are needed. Calculations have been performed using field matching methods~\cite{Zhang15}, and time domain simulations~\cite{Cho13}-\cite{Zagorodnov15}.   
In Ref.~\cite{surface_imped_flat_geometry}, for the case of flat geometry and assuming the impedance can be characterized by a surface impedance, equations for the {\it generalized} wakefields, valid for arbitrary bunch length, are derived.  By ``generalized" we here mean (point charge) wake functions for which the transverse positions of driving and test particles are arbitrary, and are not limited to being near the symmetry plane. 

In the perturbation regime, where the corrugation parameters are small compared to the aperture, the (point charge) wakes are dominated by a single mode. Thus, the longitudinal wake starts with a zero slope and, for short distances, can be approximated by a constant; for the transverse wakes, the approximation is a straight line starting from zero. However, for the RadiaBeam/LCLS dechirper we are not in the perturbation regime; consequently, the contribution of higher modes is not negligible, and the longitudinal wake will begin with an exponential drop---a ``droop," and the transverse wakes will have a droop in the slope (see {\it e.g.}~\cite{Novo15}).
In Ref.~\cite{Bane16} the generalized impedances found in Ref.~\cite{surface_imped_flat_geometry} were used to obtain analytical formulas for the generalized wakes in the ``zeroth order approximation", {\it i.e.} assuming the longitudinal wake is a constant and the transverse wakes have a constant slope.
These results give the structure of the wakes, and---for sufficiently short bunches---reasonable approximations to their strengths. 
In the LCLS (at the end of the linac) the bunch is short, with an approximately uniform distribution and a full bunch length $\lesssim 50$~$\mu$m. In Ref.~\cite{Bane16} it was also shown that, when the full bunch length is {\it e.g.} 30~$\mu$m, the zeroth order (on-axis) wakes agree to within 20--30\% with numerical results. The goal of the present report is to improve on this, by generating more accurate ``first order" expressions of the wakes.

In this report we begin our calculations with the generalized wakes obtained assuming the validity of the surface impedance model in flat geometry~\cite{surface_imped_flat_geometry}. We will limit ourselves to considering the wakes of ``pencil" beams, {\it i.e.} beams of little transverse extent, where driving and test charges follow along the same path or very close by.
We make an assumption about the form of the wakes that we check by comparing with two numerical methods: (1)~we numerically perform the inverse Fourier transform of generalized expressions for the impedance found in~\cite{surface_imped_flat_geometry} to obtain the (point charge) wake, and (2)~we obtain the bunch wake for a short Gaussian bunch using the time-domain, wakefield solving program for rectangular geometry, ECHO(2D)~\cite{Zagorodnov15}. 
There is interest in using a dechirper as a fast kicker, by passing the beam close to one jaw of the dechirper (see {\it e.g.}~\cite{Novo15}) and, in fact, this idea has already been used in two-color, self-seeding operation at the LCLS~\cite{Alberto}. Therefore, in addition to performing calculations for a beam on axis, we also study the wakes for a beam off axis.

This report is organized as follows: The first section concerns the case of a dechirper in round geometry. This is followed by the derivation of analytical expressions for longitudinal wakes in flat geometry, both on and off axis. Next come the quad and dipole wakes on axis. And finally we calculate the dipole and quad wakes away from the axis. This is followed by a short section estimating some wake effects for example LCLS machine and beam parameters. The final section gives a discussion and the conclusions.

\begin{figure}[htb]
\centering
\includegraphics[width=0.5\textwidth, trim=0mm 0mm 0mm 0mm, clip]{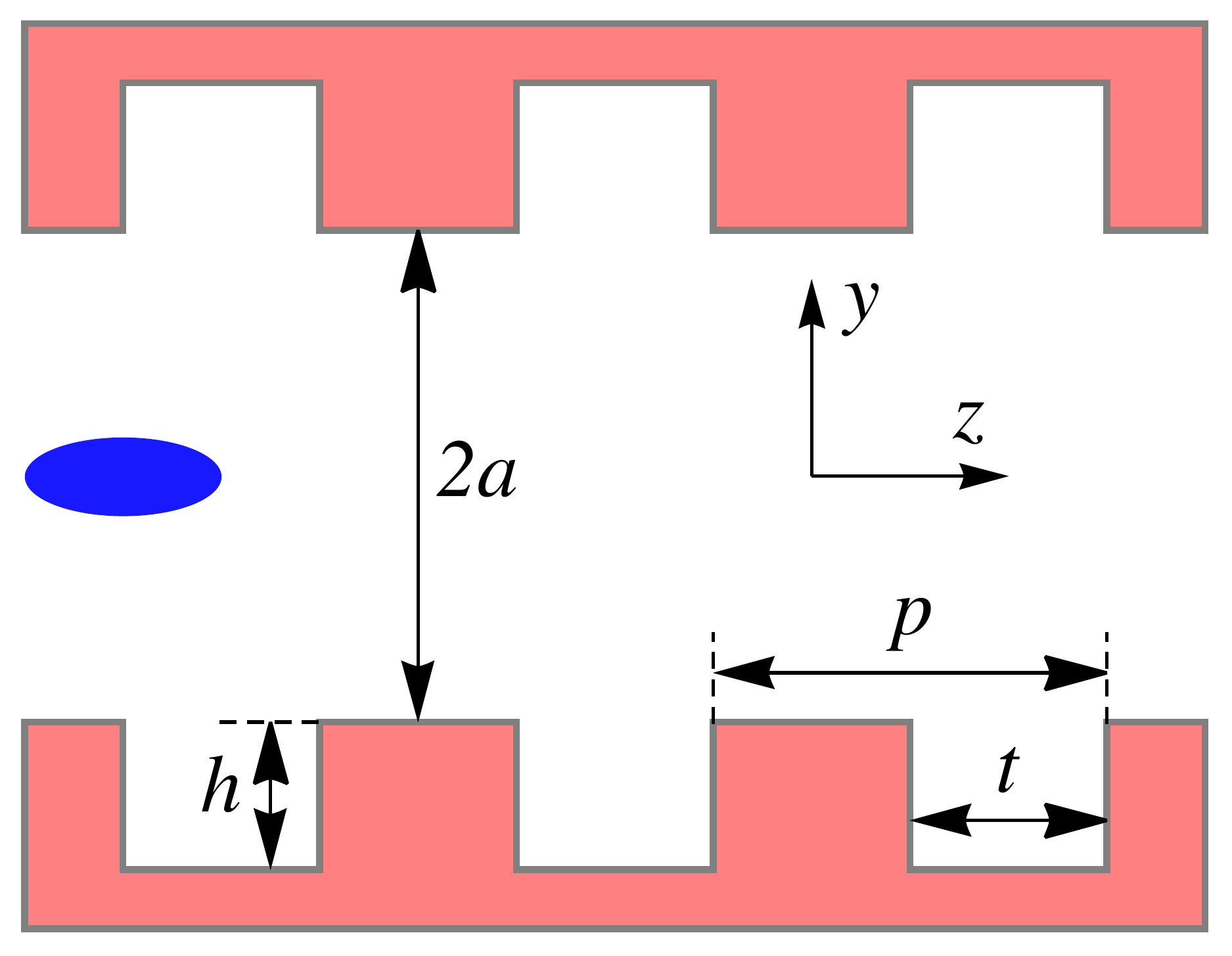}
\caption{Geometry of a dechirper unit. For a round dechirper the radius of the aperture is designated as $a$; in the flat case, for a vertical dechiper unit, $2a$ represents the gap, and the structure is unbounded in $x$. The blue ellipse represents an electron beam propagating along the $z$ axis.}
\label{fig:2}
\end{figure}

The RadiaBeam/LCLS dechirper device consists of two modules, where the first has movable jaws that are aligned horizontally (in $x$) and is called the ``vertical dechirper," and the second one with jaws that are aligned vertically (in $y$), the ``horizontal dechirper." Each section is 2~m long, and is corrugated in the longitudinal direction (Fig.~\ref{fig:2} gives the geometry of three periods of the second section).  
The parameters of the RadiaBeam/LCLS dechirper, that we use in example calculations, are given in Table~I.
 In this note most calculations are performed in Gaussian units. To convert an impedance or wake to MKS, one merely multiplies the cgs expression by $Z_0c/4\pi$, with $Z_0=377$~$\Omega$.

\begin{table}[hbt]
   \centering
   \caption{RadiaBeam/LCLS Dechirper parameters, in [mm]. The dechirper comprises a vertical and a horizontal unit, each of which consists of two flat, corrugated plates of length $L=2$~m. The length scale factor $s_{0r}$ will be discussed in the text.}
   \begin{tabular}{||l|c||}\hline 
        {Parameter name} & {Value} \\ \hline\hline 
      Period, $p$       &0.50 \\
      Longitudinal gap, $t$       &0.25  \\
      Depth, $h$       &0.50 \\
      Nominal half aperture, $a$       &0.70 \\ 
     Plate width, $w$       &12.7  \\
    Plate length, $L$       &2000  \\ 
   Length scale factor, $s_{0r}$       &0.190  \\
         \hline \hline 
   \end{tabular}
   \label{table1_tab}
\end{table}

\subsection*{Round Dechirper}

Consider a periodic, disk-loaded structure in round geometry, with iris radius $a$, period $p$, longitudinal gap $t$ (see Fig.~\ref{fig:2}). The high frequency longitudinal impedance of this structure is given by~\cite{Gluckstern89}-\cite{Yokoya99}
\begin{equation}
Z_l(k)=\frac{4i}{kca^2}\left[1+(1+i)\frac{\alpha(t/p) p}{a}\left(\frac{\pi}{kg}\right)^{1/2}\right]^{-1}\ ,\label{imp_round_eq}
\end{equation}
with $k$ the wave number, $c$ the speed of light, and $\alpha$ is a function that can be approximated by $\alpha(x)\approx1-0.465\sqrt{x}-0.070x$. The units of longitudinal impedance is [s/m$^2$].The (point charge) wake is given by the inverse Fourier transform of the impedance, 
\begin{equation}
w_l(s)=\frac{c}{2\pi}\int_{-\infty}^{\infty}dk\,Z_l(k)e^{-iks}\ ,
\end{equation}
with $s$ the distance the test particle is behind the driving charge (it is zero for $s<0$). The dimensions of longitudinal wake are [m$^{-2}$].
The wake corresponding to Eq.~\ref{imp_round_eq} is~\cite{Bane98}
\begin{equation}
w_l(s)=\frac{4}{a^2}e^{s/s_{0r}}{\rm erfc}\left(\sqrt{s/s_{0r}}\right) \ ,\label{wake_round_eq}
\end{equation}
with the distance scale factor (round case)
\begin{equation}
s_{0r}=\frac{a^2t}{2\pi\alpha^2p^2}\ .\label{s0r_eq}
\end{equation} 
For the RadiaBeam/LCLS dechirper parameters: $p=0.50$~mm, $t=0.25$~mm, (and total depth of corrugation $h=0.5$~mm). With half-gap $a=0.7$~mm, $s_{0r}=193$~$\mu$m.
Note that for large $k$, Eq.~\ref{imp_round_eq} can be Taylor expanded and written in terms of $s_{0r}$ as
\begin{equation}
Z_l(k)\approx\frac{4i}{kca^2}\left[1-\frac{(1+i)}{\sqrt{2ks_{0r}}}\right]\ .\label{imp_round_eqb}
\end{equation}

\begin{figure}[htb]
\centering
\includegraphics[width=0.6\textwidth, trim=0mm 0mm 0mm 0mm, clip]{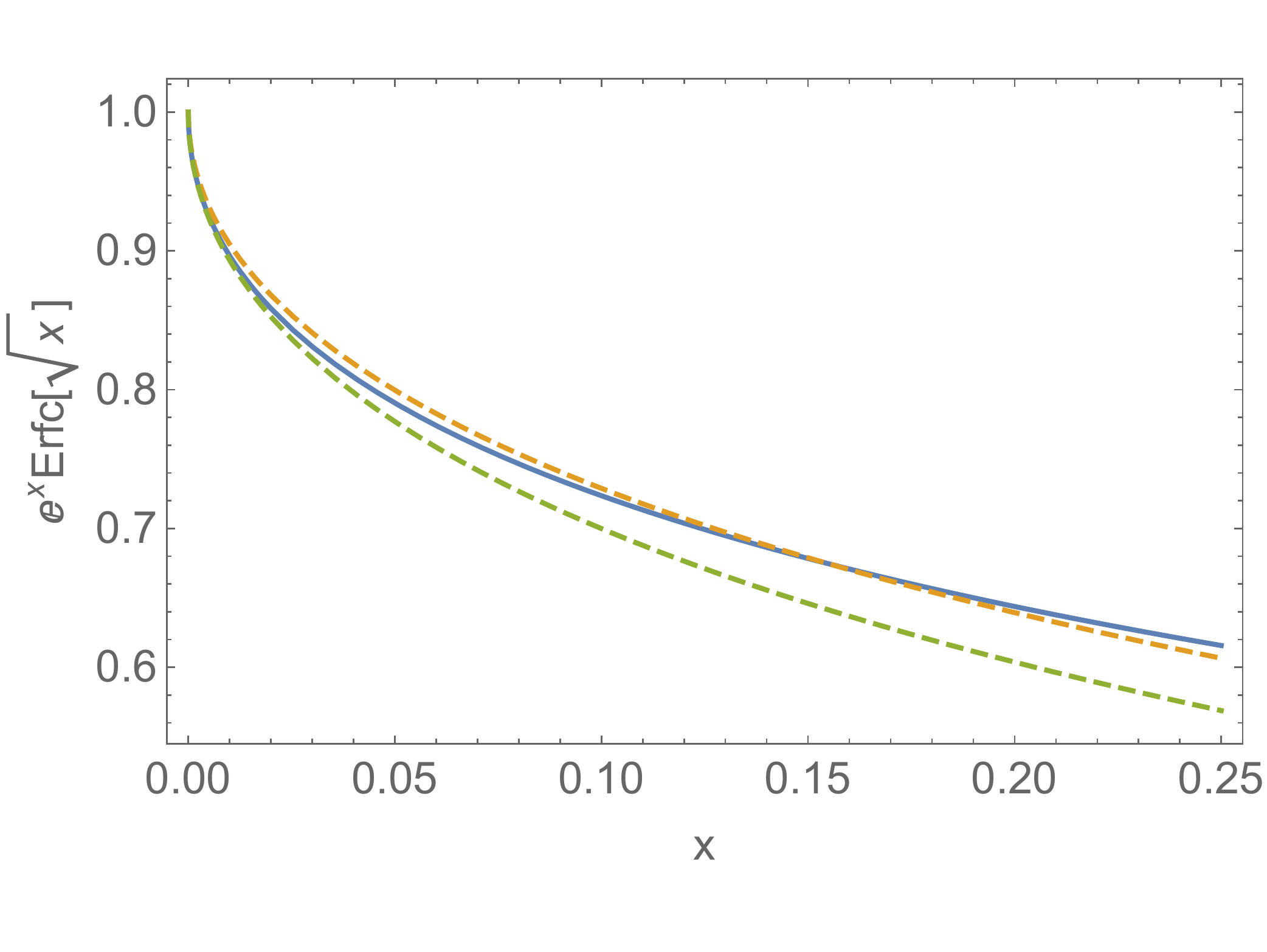}
\caption{The function $e^x{\rm erfc}(\sqrt{x})$ (blue solid), and the approximations: $e^{-\sqrt{x}}$, which  differs by less than 1.2\% over the range plotted (orange dashes); $e^{-\sqrt{4x/\pi}}$, which has the same two term Taylor expansion at the origin, $1-2\sqrt{x/\pi}$ (green dashes).}
\label{fig:1}
\end{figure}

For $s$ small compared to $s_{0r}$, Eq.~\ref{wake_round_eq} can be approximated by
\begin{equation}
w_l(s)\approx\frac{4}{a^2}e^{-\sqrt{s/s_{0r}}} \ .\label{wake_round_eqb}
\end{equation}
Note that this functional form has been found to be useful as a fitting function to represent the longitudinal wake of periodic disk-loaded accelerating structures, even for bunches that are not ultra short (see {\it e.g.} Ref.~\cite{Bane98}), and that it is used in fitting expressions for dechirper wakes provided in Ref.~\cite{Novo15}.
In Fig.~\ref{fig:1} we plot Eq.~\ref{wake_round_eq} (blue curve) and compare with Eq.~\ref{wake_round_eqb} (orange dashes) and see that the agreement is better than 1.5\% up to $s=0.25s_{0r}$. Also plotted in the figure is Eq.~\ref{wake_round_eqb}, with the scale factor changed to $\pi s_{0r}/4$,  which is a function that has the same two-term Taylor series expansion as the original equation, but does not agree as well with the real wake over the range plotted (green dashes). 

Note that the equation for the impedance, Eq.~\ref{imp_round_eq} can be written in terms of a surface impedance $\zeta(k)$ as
\begin{equation}
Z_l(k)=\frac{2}{ca}\frac{1}{(1/\zeta(k)-ika/2)}\ ,\label{Zimpb_eq}
\end{equation}
with the (high frequency) surface impedance of the structure given by
\begin{equation}
\zeta(k)=\frac{2}{a}\left(\frac{s_{0r}}{-ik}\right)^{1/2}=\frac{1}{\alpha p}\left(\frac{2t}{-i\pi}\right)^{1/2}\ .\label{zeta_eq}
\end{equation}
Note there is no dependence on the aperture, as should be the case for a surface impedance.
In terms of $s_{0r}$, Eq.~\ref{Zimpb_eq} becomes
\begin{equation}
Z_l(k)=\frac{4i}{kca^2}\left[1+\frac{(1+i)}{\sqrt{2ks_{0r}}}\right]^{-1}\ .\label{imp_roundc_eq}
\end{equation}

In the calculations for the dechirper in flat geometry below, we will make the assumption that a longitudinal impedance with the two-term, high frequency expansion of Eq.~\ref{imp_round_eqb} implies the short-range wake of Eq.~\ref{wake_round_eqb}. At the end we will test this assumption by two numerical calculation of the wake. In the transverse it is $(-ik)$ times the impedance that follows this pattern.

\subsection*{Flat Dechirper}

\subsubsection*{Longitudinal Wake On and Off Axis}

In Ref.~\cite{Bane15} the {\it generalized} longitudinal impedance is given for a flat structure whose boundary interaction can be described by a surface impedance. By generalized, we here mean to indicate the case where driving and test particle's transverse coordinates can be located anywhere within the aperture of the structure. In the present report we limit ourselves to the case where $x_0=x$ and $y_0=y$, where the driving (test) particle has subscript zero (no subscript). 
In this particular case, for a vertical dechiper unit, Eqs.~13, 14, of~\cite{Bane15} become 
\begin{equation}
 Z_l(k,y)=\frac{2\zeta}{c}\int_{-\infty}^\infty dq\,q\,\mathrm{csch}^3(2qa) f(q,y)\label{tildeZl2_eq}\ ,
\end{equation}
where the function $f=N/D$, with
\begin{eqnarray}\label{eq:9-1}
N&=&q\big(\cosh[2q(a-y)]-2+\cosh[2q(a+y)]\big)
-ik\zeta\big(\sinh[2q(a-y)]\nonumber\\
&&+\sinh[2q(a+y)]\big)\ \nonumber\\
D&=&\big[q\,\mathrm{sech}(qa)-ik\zeta\mathrm{csch}(qa)][q\,\mathrm{csch}(qa)-ik\zeta\mathrm{sech}(qa)\big]\ .
\end{eqnarray}
Substituting Eq.~\ref{zeta_eq} into Eq.~\ref{tildeZl2_eq} and numerically performing the integration, we can obtain the impedance; then inverse Fourier transforming, we obtain the wake.

In Ref.~\cite{Bane16} the wake at the origin ($s=0^+$) was obtained by expanding the impedance at high $k$, keeping only the leading order ($1/k$) term in the impedance, and then inverse Fourier transforming. This method gave an upper bound to the short-range wake function. Here we expand again, but keep also the next term ($1/k^{3/2}$), and then assume that the dependence, $W_l\sim e^{-\sqrt{s/s_0}}$, is a good approximation to the short-range wake. Thus, this calculation, in addition to giving us the amplitude at the origin, also gives us the distance scale factor $s_0$. At the end we will need to verify that this functional dependence is indeed a good approximation to the short-range wake.

Substituting for $\zeta(k)$ (Eq.~\ref{zeta_eq}) into the impedance, then expanding in $1/k$ and keeping the first two terms, we obtain
\begin{eqnarray}
 Z_l(k,y)&=&\frac{1}{kc}\int_{-\infty}^\infty dq\Big({2iq}\cosh(2qy){\rm csch}(2qa)\\
&+&\frac{q^2a\,{\rm csch}^3(2qa)}{2(1+i) \sqrt{2s_{0r}k}}\left[4\sinh(2qa)+\sinh(2q[2a+y])+\sinh(2q[2a-y])\right]\Big)\nonumber\ .
\end{eqnarray}
The integrals can be performed analytically, giving
\begin{equation}
Z_l(k,y)=\frac{\pi^2i}{4kca^2}\sec^2(\frac{\pi y}{2a})\left(1-\frac{(1+i)}{\sqrt{8ks_{0r}}}\left[1+\frac{1}{3}\cos^2(\frac{\pi y}{2a})+(\frac{\pi y}{2a})\tan(\frac{\pi y}{2a})\right]\right)\ .
\end{equation}
Comparing this expansion with that for the round case (Eq.~\ref{imp_round_eqb}), we see that the equivalent distance scale factor for the flat case, $s_{0l}$, is given by
\begin{equation}
s_{0l}=4s_{0r}\left[1+\frac{1}{3}\cos^2(\frac{\pi y}{2a})+(\frac{\pi y}{2a})\tan(\frac{\pi y}{2a})\right]^{-2}\ .\label{s0l_eq}
\end{equation}
And the short-range longitudinal wake is given as
\begin{equation}
w_l(s,y)=\frac{\pi ^2}{ 4a^2}\sec^2\left(\frac{\pi y}{2a}\right)e^{-\sqrt{s/s_{0l}}}\ .\label{Wl_eq}
\end{equation}
Note that for the beam on axis, $s_{0l}=\frac{9}{4}s_{0r}$.

\begin{figure}[htb]
\centering
\includegraphics[width=0.6\textwidth, trim=0mm 0mm 0mm 0mm, clip]{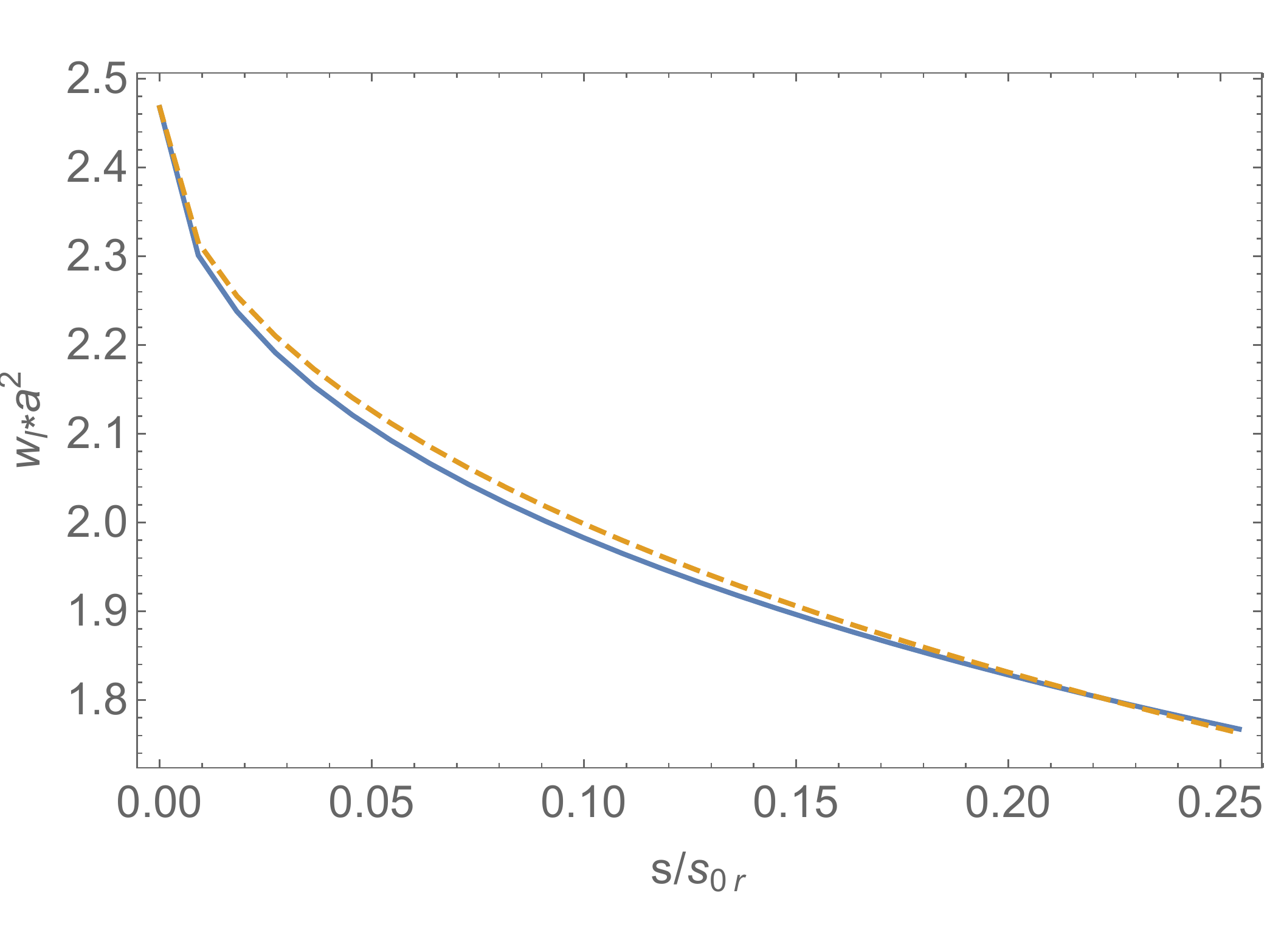}
\caption{Longitudinal wake on axis for the RadiaBeam/LCLS dechirper nominal parameters, comparing the numerical inverse Fourier transform of the general impedance equation, Eq.~\ref{tildeZl2_eq} (blue curve), with the approximation, Eq.~\ref{Wl_eq} (the dashes). The scale factor $s_{0r}=190$~$\mu$m.}
\label{Wlb_fi}
\end{figure}

Using the RadiaBeam/LCLS dechirper nominal parameters (Table~1), for the case with the beam on axis, we compare $w_l(s,y=0)$ from Eq.~\ref{Wl_eq} with the numerical result obtained by taking the inverse Fourier transform of the general impedance equation, Eq.~\ref{tildeZl2_eq}, with no approximations. The results are shown in Fig.~\ref{Wlb_fi}, where the numerical results are given by the blue curve, and the analytical approximation by the dashed orange curve. We see good agreement.  

We also compare with results using the time domain, wakefield solver for structures with rectangular geometry, ECHO(2D)~\cite{Zagorodnov15}. The earlier calculations were for flat geometry, which means---for a vertical dechirper---parallel plates that extend to infinity in both horizontal directions, and have corrugations in $y$ {\it vs.} $z$. ECHO(2D), however, assumes smooth side walls at $x=\pm w/2$, and the wakes come as a sum of discrete modes, with odd mode numbers $m$, corresponding to horizontal mode wave numbers $k_x=m\pi/w$. If the aspect ratio $2a/w$ is small enough, and a sufficient number of modes are summed, then the flat, short-range wake result and the ECHO(2D) result should agree. For our ECHO(2D) caculations we take $a=0.7$~mm and $w=12$~mm, and the aspect ratio $2a/w=0.12$ is sufficiently small. The highest mode number in the calculations is $m=89$; such a large number was needed for good convergence for the off-axis examples discussed below. For the ECHO runs we simulated a Gaussian driving bunch with rms length $\sigma_z=10$~$\mu$m passing through $L=2$~m of structure (for the wakes, we then normalize to length). Note that since the catch-up distance $z_{cu}=a^2/2\sigma_z=2.4$~cm is short compared to the structure length, the transient wake contribution is negligible and can be ignored.

\begin{figure}[htb]
\centering
\includegraphics[width=0.6\textwidth, trim=0mm 0mm 0mm 0mm, clip]{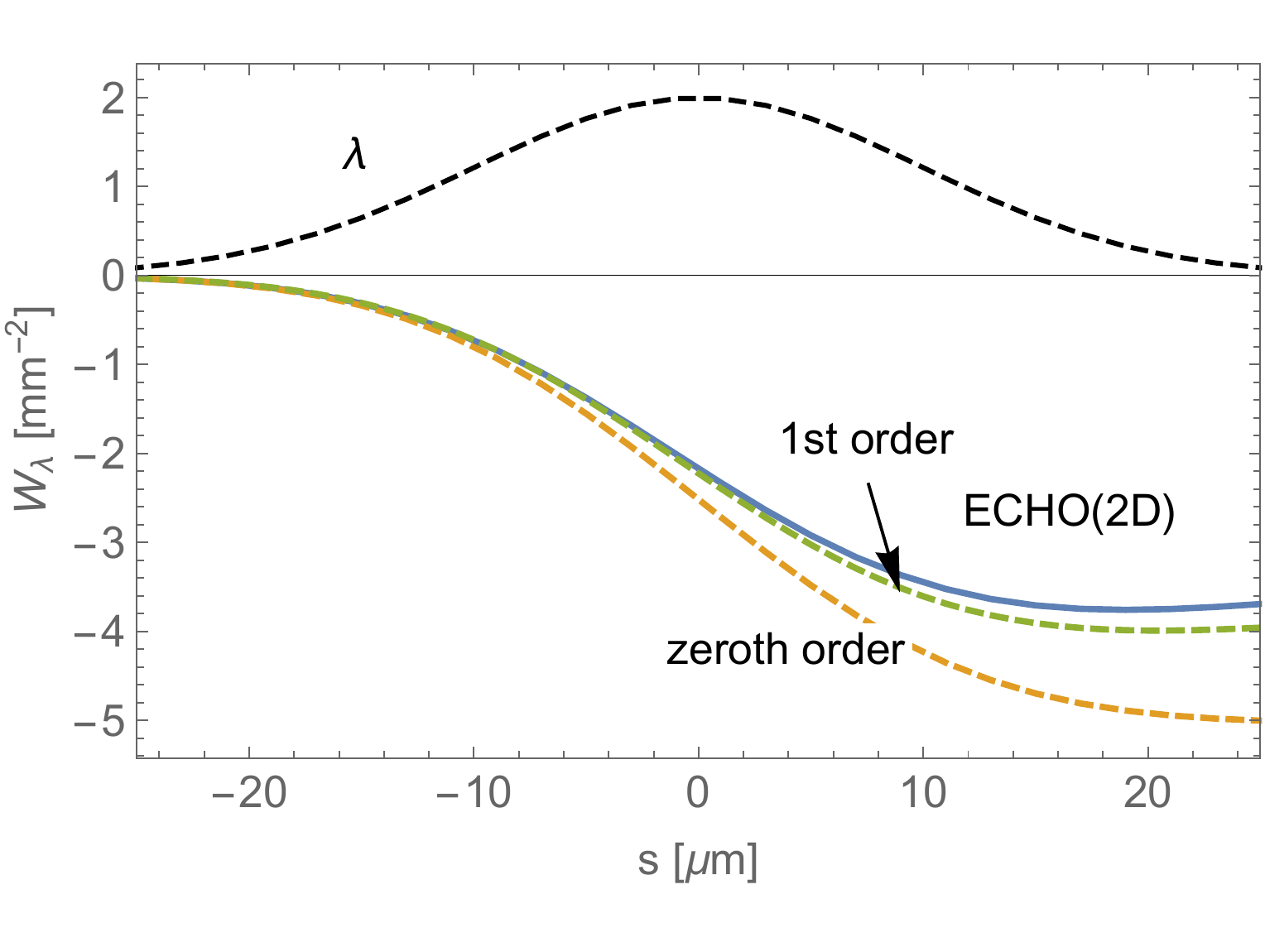}
\caption{Longitudinal bunch wake for a Gaussian beam on axis, with $\sigma_z=10$~$\mu$m, in a 2-m section of the RadiaBeam/LCLS dechirper. Given are the numerical results of ECHO(2D) (blue), and the analytical zeroth order (red) and 1st order results (green). The bunch shape $\lambda(s)$ with the head to the left is shown in black.}
\label{Wl_fi}
\end{figure}

 For the analytical wake we need to convolve with the longitudinal bunch shape $\lambda(s)$:
\begin{equation}
W_\lambda(s)=\int_0^\infty ds'\,w_l(s')\lambda(s-s')\ .
\end{equation}
In Fig.~\ref{Wl_fi} we show the bunch wake $W_\lambda(s)$ according to the ECHO results (blue), the analytical zeroth order result (orange), and the result using the wake of Eq.~\ref{Wl_eq} (green, a result we call ``first order"). We see that the first order result agrees well with that of ECHO; in addition, we see that it is a great improvement over the zeroth order result. Note, however, that the agreement with ECHO is not perfect, while it seemed near-perfect in the comparison to the numerical impedance calculation (Fig.~\ref{Wlb_fi}). This suggests that there is some inaccuracy in the impedance approach, 
 probably related to inaccuracy in the high frequency surface impedance used to represent the structure.

\begin{figure}[htb]
\centering
\includegraphics[width=0.6\textwidth, trim=0mm 0mm 0mm 0mm, clip]{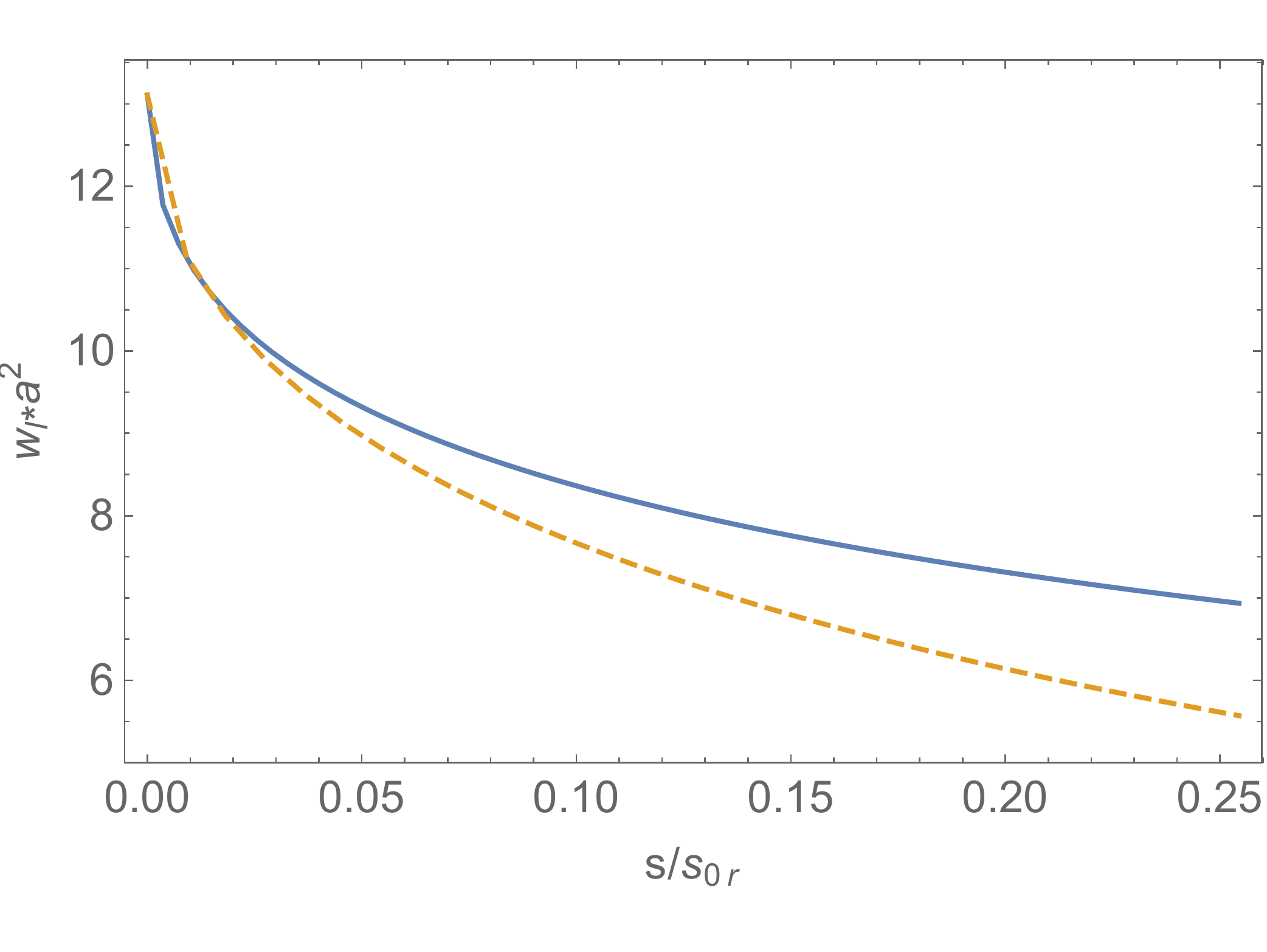}
\caption{Longitudinal wake off axis by $y=0.5$~mm, comparing the numerical inverse Fourier transform of the general impedance equation, Eq.~\ref{tildeZl2_eq} (blue curve), with the first order approximation, Eq.~\ref{Wl_eq} (the dashes). 
}
\label{Wlb_yoff_fi}
\end{figure}

\begin{figure}[htb]
\centering
\includegraphics[width=0.6\textwidth, trim=0mm 0mm 0mm 0mm, clip]{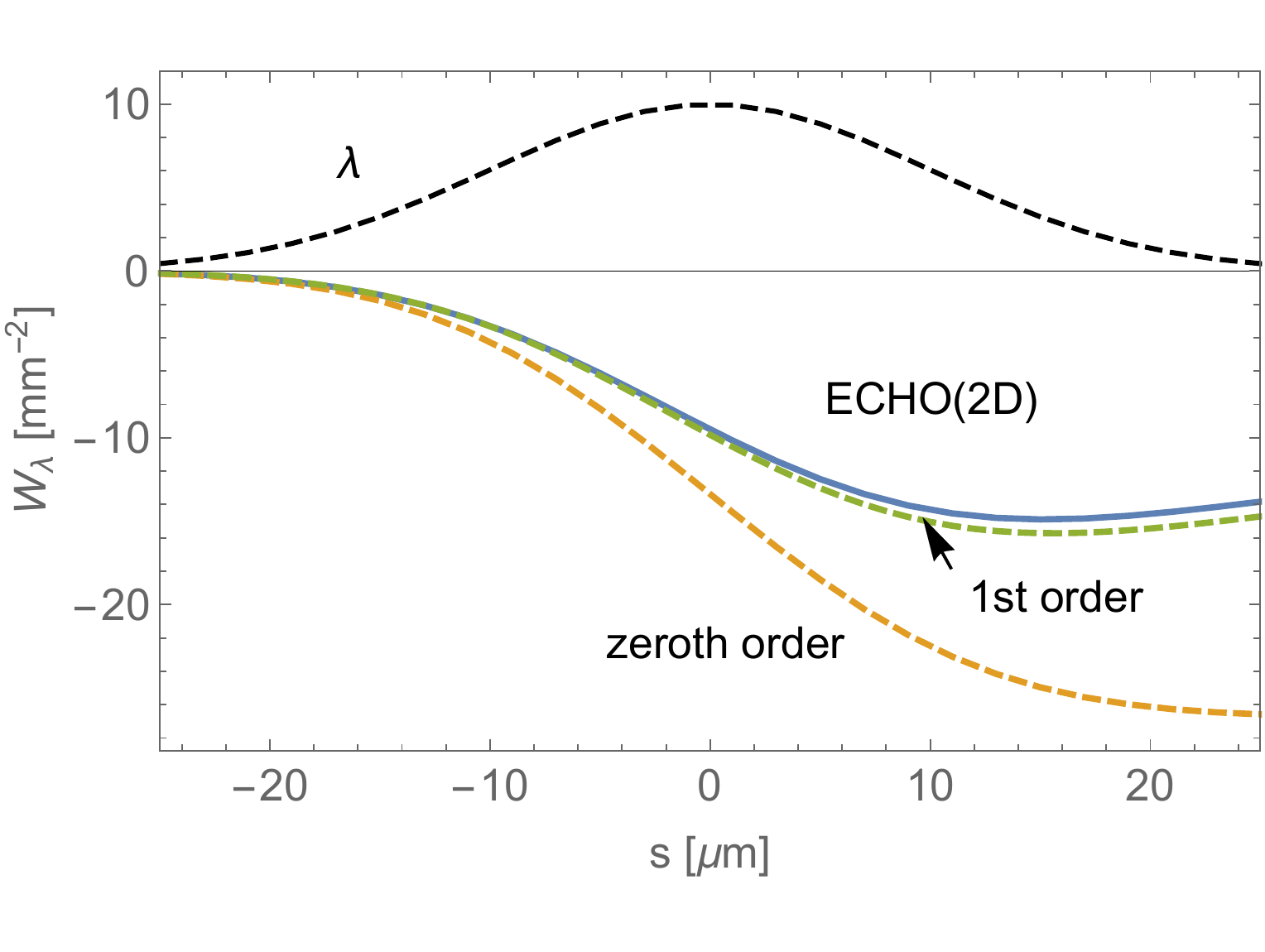}
\caption{Longitudinal bunch wake for a Gaussian beam with $\sigma_z=10$~$\mu$m offset in the RadiaBeam/LCLS dechirper. The offset is $y=0.5$~mm and the half-aperture is the nominal $a=0.7$~mm. Given are the numerical results of ECHO(2D) (blue), and the analytical zeroth order (red) and 1st order results (green). The bunch shape $\lambda(s)$ with the head to the left is shown in black.}
\label{Wl_yoff_fi}
\end{figure}

In Figs.~\ref{Wlb_yoff_fi} and \ref{Wl_yoff_fi} the comparison calculations are repeated for a beam offset by $y=0.5$~mm. We see that the analytical result for $w_l$ (``first order") does not agree so well with the numerical result for $s\sim 0.25s_{0r}=50$~$\mu$m. However, it still agrees well with the Gaussian wake obtained by ECHO. This is because, in the former case, the agreement is still good until $s=0.05s_{0r}=10$~$\mu$m. The implication is that, for longer bunch lengths, the off-axis wake will begin to deviate from the ECHO results, too. Note that the first order result is a great improvement over the zeroth order one. 

\subsubsection*{Transverse Wakes Near the Axis}

Near the axis of a flat, vertical dechirper unit, for a complete description of the transverse impedance we require two independent functions, the vertical quad and dipole impedances, $Z_{yq}$ and $Z_{yd}$.
The transverse impedances in $y$ and $x$ are then given, to leading order in offset, by (see {\it e.g.} \cite{Bane15})
\begin{equation}
\tilde Z_y=y_0Z_{yd}+yZ_{yq}\ , \quad\quad\quad\quad \tilde Z_x=(x_0-x)Z_{yq}\ ,\label{qd_eq}
\end{equation}
where ($x_0$,$y_0$) and  ($x$,$y$) are the offsets of, respectively, the driving and test particles. Note that the transverse wakes have the same properties, with the two wake functions being the vertical quad and dipole wakes, $w_{yq}$ and $w_{yd}$.

The transverse wakes, like the quad wake, are obtained from the impedance by taking the inverse Fourier transform:
\begin{equation}
w_{yq}(s)=\frac{ic}{2\pi}\int_{-\infty}^\infty dk\, Z_{yq}(k)e^{-i k s}\ .
\end{equation}
In the transverse case, for example with the quad wake, it is the slope of the wake for which $w'_{yq}(s)=Ae^{-\sqrt{s/s_{0q}}}$, with $A$ and $s_{0q}$ constants. Thus, by expanding $-ikZ_{yq}(k)$ to second order in $1/k$ we will find the short-distance approximation of $w'_{yq}(s)$. And then the wake itself will be given by integrating; {\it e.g.} $w_{yq}=2As_{yq}[1-(1+\sqrt{s/s_{0q}})e^{-\sqrt{s/s_{0q}}}]$.

The vertical quad impedance is given by~\cite{Bane15} 
\begin{equation}
Z_{yq}(k)=
     \frac{2}{kca^3}
     \int_{0}^\infty
     dx\,x^2
     \frac{
     \mathrm{sech}(x)
     }
     {
     \cosh(x)/\zeta
     -
     ika\sinh(x)/x
     }\ .\label{Zq_eq}
\end{equation}
Multiplying by $(-ik)$ and expanding in terms of $1/k$ for high $k$, we obtain
\begin{equation}
 -ikZ_{yq}(k)=\frac{1}{kca^3}\int_{0}^\infty dx\left[2x^3\mathrm{csch}(x)\mathrm{sech}(x)
-\frac{(1+i)}{ \sqrt{2s_{0r}k}}x^4\mathrm{csch^2(x)}\right]\ .
\end{equation}
The integrals are performed analytically, and after rearrangement we obtain
\begin{equation}
 -ikZ_{yq}(k)=\frac{i\pi^4}{32kca^4}\left[1-\frac{(1+i)}{\sqrt{2ks_{0r}}}\left(\frac{16}{15}\right)\right]\ .
\end{equation}
Thus, the quad wake is given by 
\begin{equation}
w_{yq}(s)\approx
     \frac{\pi^4}{16a^4}
   s_{0q}\left[1-\left(1+\sqrt{\frac{s}{s_{0q}}}\right)e^{-\sqrt{s/s_{0q}}}\right]
    \ ,\label{wqd2_eq}
\end{equation}
with
\begin{equation}
s_{0q}=s_{0r}\left(\frac{15}{16}\right)^2\ .\label{s0q_eq}
\end{equation}

Finding the vertical dipole wake follows a similar process. The vertical dipole impedance is given by~\cite{Bane15} 
\begin{equation}
Z_{yd}(k)= 
     \frac{2}{kca^3}
     \int_{0}^\infty
     dx\,x^2
     \frac{
     \mathrm{csch}(x)
     }
     {
     \sinh(x)/\zeta
     -
     ika\cosh(x)/x
     }\ .\label{Zd_eq}
\end{equation}
Multiplying by $(-ik)$ and expanding in terms of $1/k$ for high $k$, we obtain
\begin{equation}
 -ikZ_{yq}(k)=\frac{1}{kca^3}\int_{0}^\infty dx\left[2x^3\mathrm{csch}(x)\mathrm{sech}(x)
-\frac{(1+i)}{ \sqrt{2s_{0r}k}}x^4\mathrm{sech^2(x)}\right]\ .
\end{equation}
The integrals are performed analytically, and after rearrangement we obtain
\begin{equation}
 -ikZ_{yq}(k)=\frac{i\pi^4}{32kca^4}\left[1-\frac{(1+i)}{\sqrt{2ks_{0r}}}\left(\frac{14}{15}\right)\right]\ .
\end{equation}
 The form of the dipole wake is the same as the quad wake (Eq.~\ref{wqd2_eq}), but with the quad scale factor replaced by the dipole scale factor,
\begin{equation}
s_{0d}=s_{0r}\left(\frac{15}{14}\right)^2\ .\label{s0d_eq}
\end{equation}

\begin{figure}[htb]
\centering
\includegraphics[width=0.6\textwidth, trim=0mm 0mm 0mm 0mm, clip]{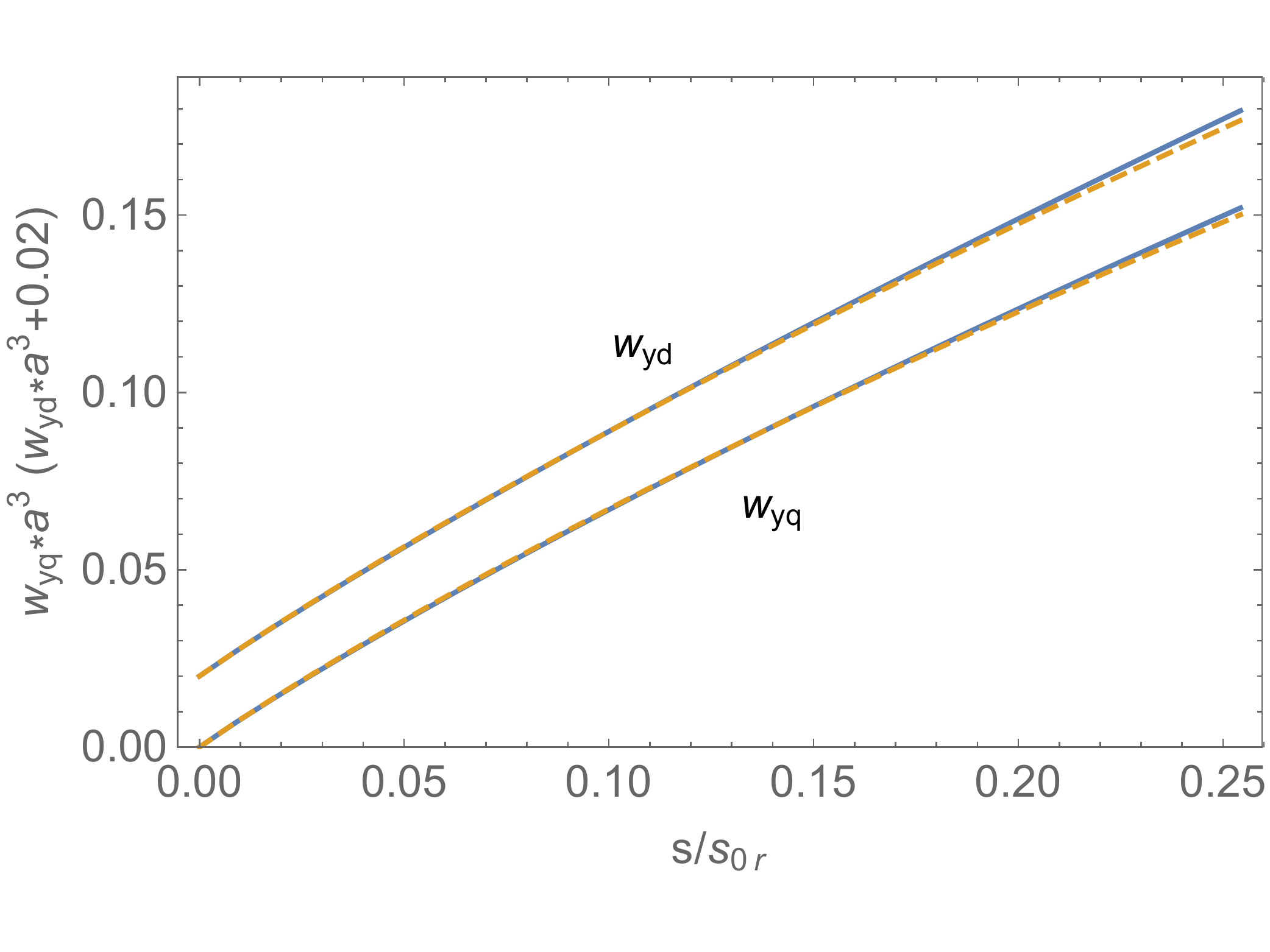}
\caption{Quad and dipole wakes on axis for the RadiaBeam/LCLS dechirper nominal parameters, comparing the numerical inverse Fourier transform of the impedance equations, Eqs.~\ref{Zq_eq}, \ref{Zd_eq} (blue curves), with the approximations, Eqs.~\ref{wqd2_eq}, \ref{s0q_eq}, \ref{s0d_eq} (the dashes). Note that the curves for $w_{yd}a^3$ have been shifted up by 0.02 units to improve visibility. 
}
\label{Wqdb_fi}
\end{figure}

\begin{figure}[htb]
\centering
\includegraphics[width=0.6\textwidth, trim=0mm 0mm 0mm 0mm, clip]{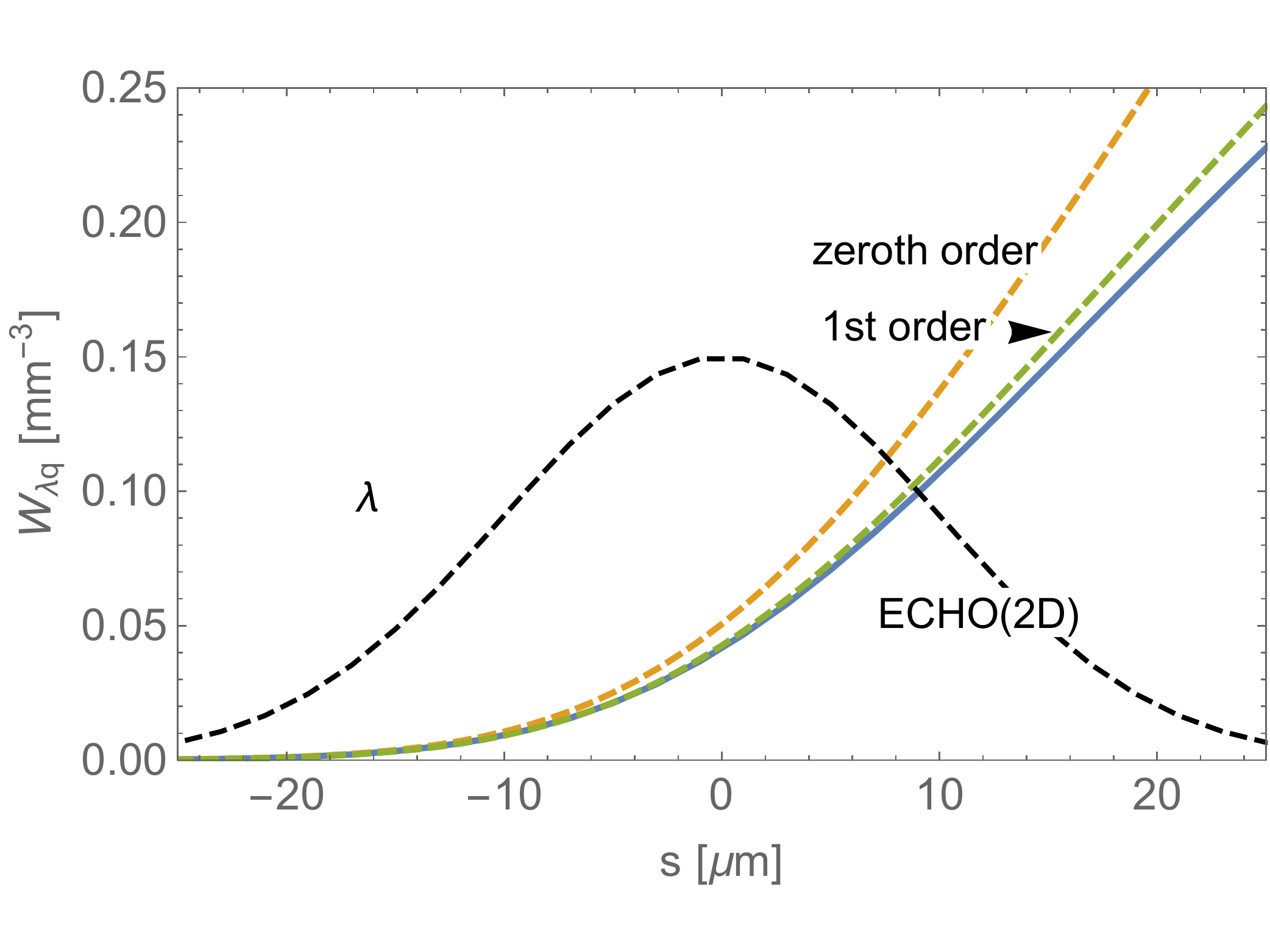}
\caption{Quad bunch wake for a Gaussian beam on axis, with $\sigma_z=10$~$\mu$m, in the RadiaBeam/LCLS dechirper. Given are the numerical results of ECHO(2D) (blue), and the analytical zeroth order (red) and 1st order results (green). The bunch shape $\lambda(s)$ with the head to the left is shown in black.}
\label{Wq_fi}
\end{figure}

\begin{figure}[htb]
\centering
\includegraphics[width=0.6\textwidth, trim=0mm 0mm 0mm 0mm, clip]{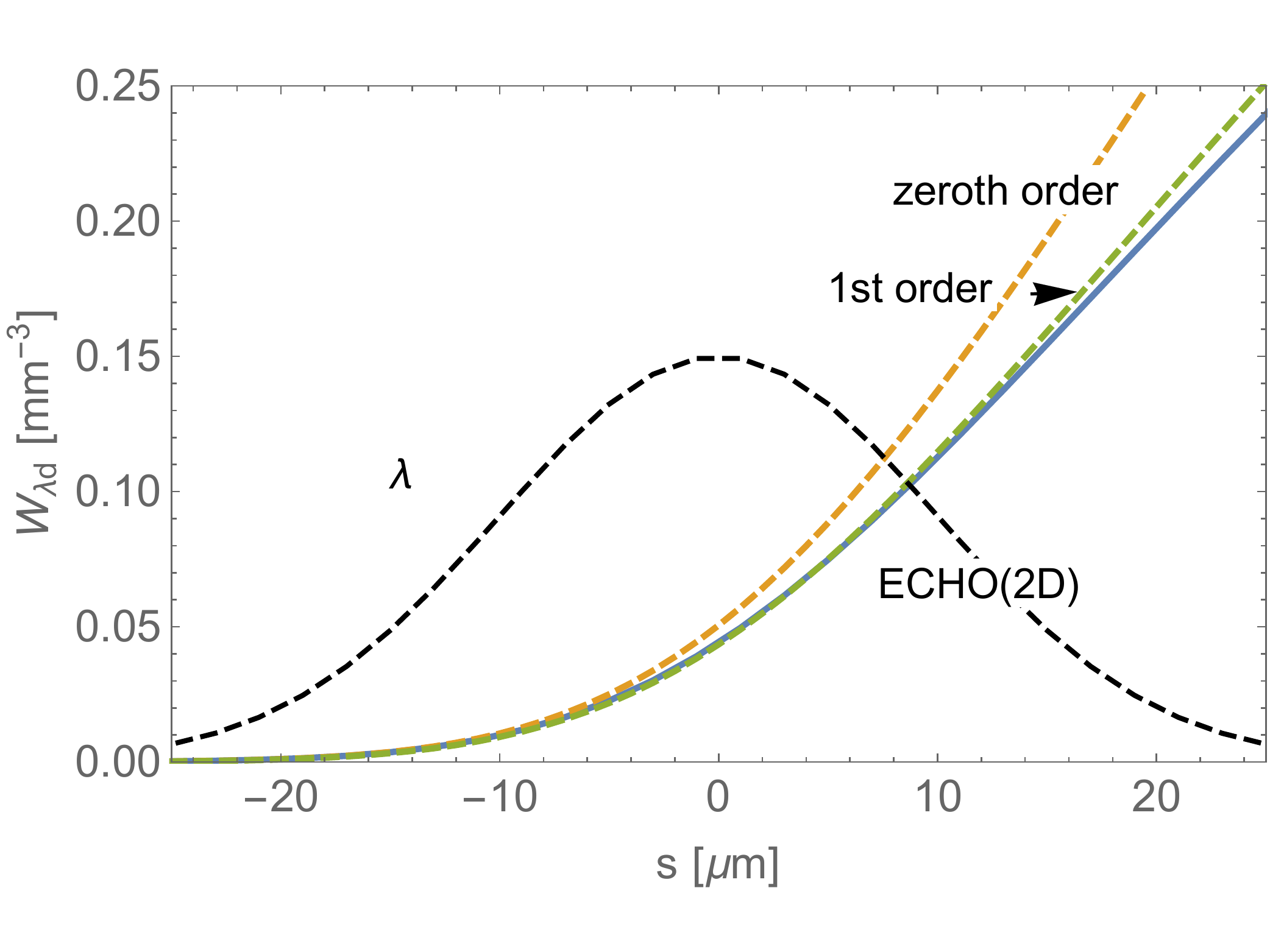}
\caption{Dipole bunch wake for a Gaussian beam on axis, with $\sigma_z=10$~$\mu$m, in the RadiaBeam/LCLS dechirper. Given are the numerical results of ECHO(2D) (blue), and the analytical zeroth order (red) and 1st order results (green). The bunch shape $\lambda(s)$ with the head to the left is shown in black.}
\label{Wd_fi}
\end{figure}

For the RadiaBeam/LCLS dechirper nominal parameters, for the quad and dipole wakes, we compared the numerical inverse Fourier transform of the impedance equations, Eqs.~\ref{Zq_eq}, \ref{Zd_eq}, with the approximations, Eqs.~\ref{wqd2_eq}, \ref{s0q_eq}, \ref{s0d_eq}. The results are shown in Fig.~\ref{Wqdb_fi}, with the numerical results given in blue, and the analytical ones by red dashes. (Note that the curves for $w_{yd}a^3$ have been shifted up in the plot, by 0.02 units, to improve the visibility.) We see that the agreement in both cases is very good. Finally, in Figs.~\ref{Wq_fi}, \ref{Wd_fi}, we compare the analytical quad and dipole bunch wakes for a $\sigma_z=10$~$\mu$m Gaussian beam in the RadiaBeam/LCLS dechirper with ECHO(2D) numerical calculations. Again we see that the agreement with the analytical model (``first order") is very good.

\subsubsection*{Transverse Wake Away from Axis}

There is interest in using a dechirper as a fast kicker by passing the beam near to one of the jaws. For a beam passing close to one jaw, the bunch head receives no kick while the tail receives a large transverse kick; in between, for the normal uniform bunch distribution, the kick varies quadratically with distance from the head. Thus, we are interested in knowing the vertical wake kick far from the axis, and we are also interested in knowing the wake defocusing effect there. 

We can extend the concepts of the dipole and quad wakes and impedances near the axis to beams that are off axis. Consider driving and test particles that have nearly the same offset $y$, in a vertical dechirper unit. For a driving particle at $(0,y)$, and a test particle at $(x,y+\Delta y)$, with $(x,\Delta y)\ll y$, the total transverse impedances can be written to leading orders as
\begin{equation}
\tilde Z_y(y)=\tilde Z_{yd}(y)+\Delta yZ_{yq}(y)\ , \quad\quad\quad\quad \tilde Z_x(y)=-xZ_{yq}(y)\ ,\label{qdb_eq}
\end{equation}
(and similar for the corresponding wakes). The dipole impedance defined here, $\tilde Z_{yd}(y)$, is different from that defined near the axis, $Z_{yd}$ (see Eq.~\ref{qd_eq}): it is still independent of test particle deviation ($\Delta y$); however, it is a function of the nominal offset of both particles ($y$), and it is not normalized to an offset.

For a driving particle at $(x_0,y_0)$ and test particle at $(x,y)$ the generalized vertical impedance is given by~\cite{Bane15} 
\begin{equation}
\tilde Z_y(k,y)=-\frac{2\zeta}{ck}\int_{-\infty}^\infty dq\,q^2\,\mathrm{csch}^3(2qa) g(q,y) e^{-iq(x-x_0)}\ ,\label{tildeZy2_eq}
\end{equation}
where $g(q,y)=N'/D$, with
\begin{eqnarray}\label{eq:9-1b}
N'=q(\sinh[q(2a-y-y_0)]&&+2\sinh[q(y-y_0)]-\sinh[q(2a+y+y_0)])\nonumber\\
&&\!\!\!\!\!\!\!\!\!\!\!\!\!\!\!\!\!\!-ik\zeta(\cosh[q(2a-y-y_0)]-\cosh[q(2a+y+y_0)])\ ,\nonumber\\
D=[q\,\mathrm{sech}(qa)-ik\zeta&&\mathrm{csch}(qa)][q\,\mathrm{csch}(qa)-ik\zeta\mathrm{sech}(qa)]\ .
\end{eqnarray}
For the transverse dipole impedance $\tilde Z_{yd}(k,y)$ for a pencil beam, we set $x_0=x$ and $y_0=y$.
We then multiply the impedance by $(-ik)$, which gives the impedance corresponding to the slope of the vertical impedance, and substitute for $\zeta(k)$ (Eq.~\ref{zeta_eq}); then we expand in $1/k$ and keep the first two terms; we arrive at
\begin{eqnarray}
-ik \tilde Z_{yd}(k,y)&=&\frac{1}{kca^3}\int_{-\infty}^\infty dx\Big[{2x^2}\mathrm{csch}(2x)\sinh(2xy/a)\\
&-&\frac{(1+i)}{ \sqrt{2s_{0r}k}}x^3\mathrm{coth}(2x)\mathrm{csch}(2x)\sinh(2xy/a)\Big]\nonumber\ .
\end{eqnarray}
The integrals can be performed analytically, and after rearranging terms we find
\begin{eqnarray}
-ik \tilde Z_{yd}(k,y)&=&\frac{\pi^3}{8kca^3}\sec^2\left(\frac{\pi y}{2a}\right)\tan\left(\frac{\pi y}{2a}\right)\bigg(1\\
&-&\left.\frac{(1+i)}{ \sqrt{8s_{0r}k}}\left[\frac{3}{2}+\frac{\pi y}{a}\mathrm{csc}(\frac{\pi y}{a})-\frac{\pi y}{2a}\mathrm{cot}(\frac{\pi y}{a})\right]\right)\nonumber\ .
\end{eqnarray}

Comparing this expansion with that for the round case (Eq.~\ref{imp_round_eqb}), we see that the slope of the wake, $w'_{yd}(s)\sim e^{-\sqrt{s/s_{0y}}}$, with the equivalent distance scale factor given by
\begin{equation}
s_{0yd}=4s_{0r}\left[\frac{3}{2}+\frac{\pi y}{a}\mathrm{csc}(\frac{\pi y}{a})-\frac{\pi y}{2a}\mathrm{cot}(\frac{\pi y}{a})\right]^{-2}\ .\label{s0yd_eq}
\end{equation}
Integrating $w'_y$ over $s$, we find that the short-range vertical wake is given by
\begin{equation}
w_{yd}(s)\approx
     \frac{\pi^3}{4a^3}\sec^2\left(\frac{\pi y}{2a}\right)\tan\left(\frac{\pi y}{2a}\right)
   s_{0yd}\left[1-\left(1+\sqrt{\frac{s}{s_{0yd}}}\right)e^{-\sqrt{s/s_{0yd}}}\right]
    \ .\label{wqd_yoff_eq}
\end{equation}

For the quad impedance for arbitrary offset, we take $\partial \tilde Z_y(k,y)/\partial y$ using Eq.~\ref{tildeZy2_eq}, and then let $x_0=x$ and $y_0=y$ to obtain:
\begin{equation}
  Z_{yq}(k,y)=\frac{2\zeta}{ck}\int_{-\infty}^\infty dq\,q^3\,\mathrm{csch}^3(2qa) f(q,y) \ ,\label{tildeZyq2_eq}
\end{equation}
where $f(q,y)=N/D$, with $N$, $D$, defined in Eq.~\ref{eq:9-1}. 
Multiplying by $(-ik)$, expanding to second order, and integrating as before, we find that the quad wake with the beam offset from the axis is given by
\begin{equation}
w_{yq}(s)\approx
     \frac{\pi^4}{16a^4}\left[2-\cos\left(\frac{\pi y}{a} \right) \right]\sec^4\left(\frac{\pi y}{2a} \right)
   s_{0yq}\left[1-\left(1+\sqrt{\frac{s}{s_{0yq}}}\right)e^{-\sqrt{s/s_{0yq}}}\right]
    \ ,\label{wq_yoff_eq}
\end{equation}
with
\begin{equation}
s_{0yq}=4s_{0r}\left(\frac{56-\cos2\theta}{30}+\frac{\frac{3}{10}+\theta\sin2\theta}{2-\cos2\theta}+2\theta\tan\theta\right)^{-2}\ \label{s0yq_eq}
\end{equation}
and $\theta=\pi y/(2a)$. Note that for $y=0$ the result agrees with Eq.~\ref{wqd2_eq}.

\begin{figure}[htb]
\centering
\includegraphics[width=0.6\textwidth, trim=0mm 0mm 0mm 0mm, clip]{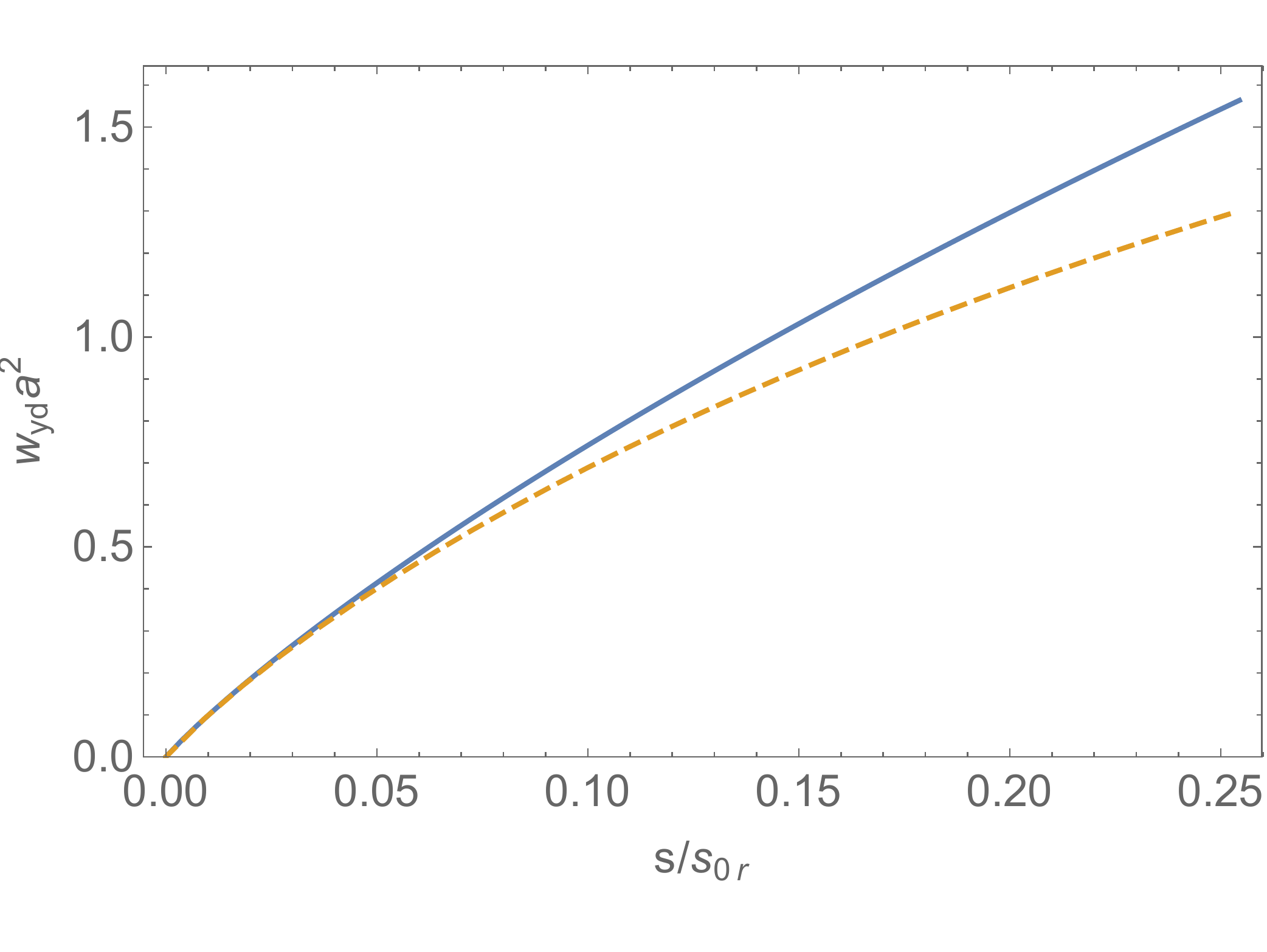}
\caption{Dipole wake at offset $y=0.5$~mm for the RadiaBeam/LCLS dechirper, comparing the numerical inverse Fourier transform of the impedance equation, Eq.~\ref{tildeZy2_eq} with $x_0=x$, $y_0=y$ (blue curve), with the approximation, Eq.~\ref{wqd_yoff_eq} (the dashes). 
}
\label{Wd_yoff_fi}
\end{figure}

\begin{figure}[htb]
\centering
\includegraphics[width=0.6\textwidth, trim=0mm 0mm 0mm 0mm, clip]{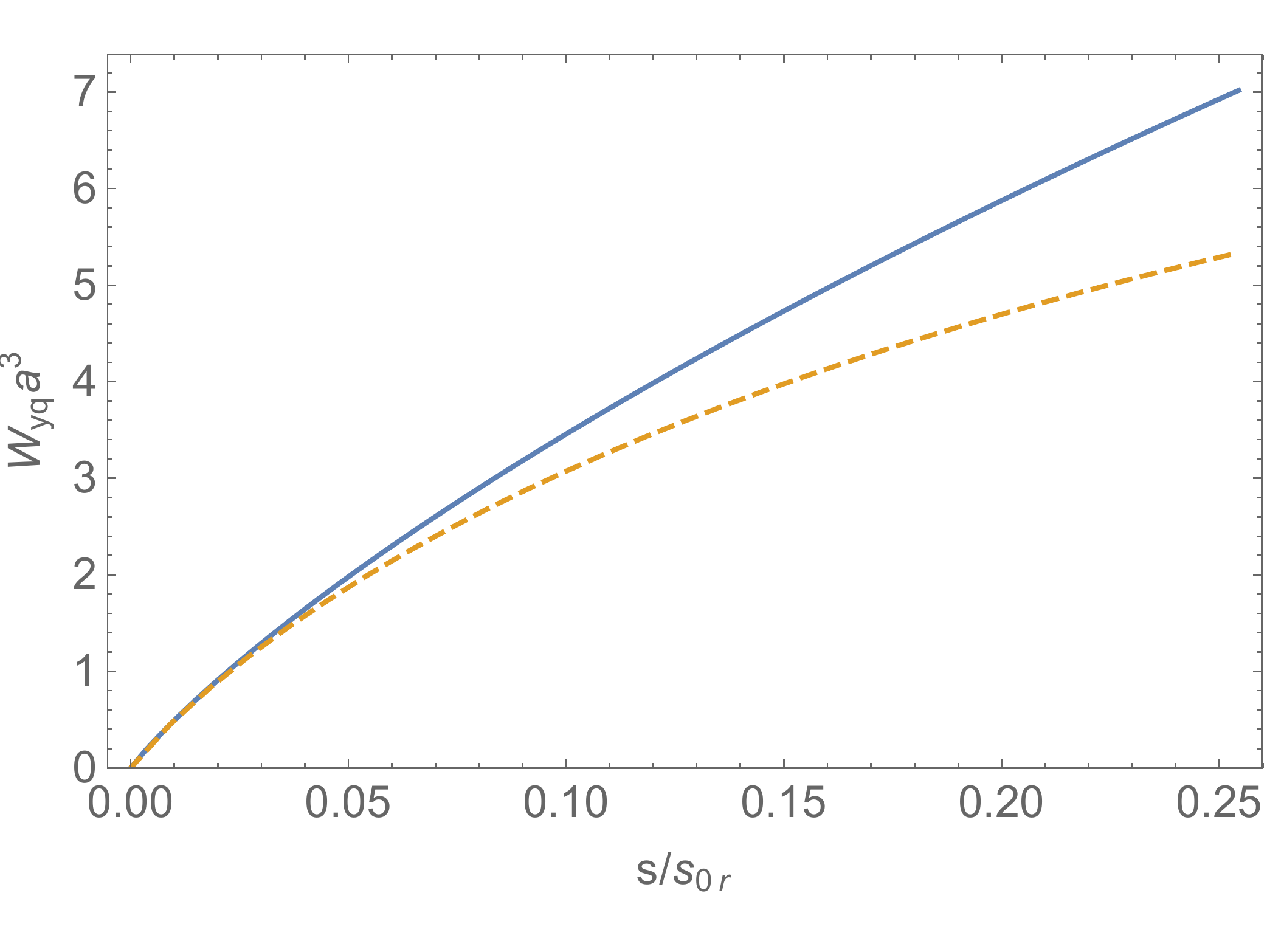}
\caption{Quad wake at offset $y=0.5$~mm for the RadiaBeam/LCLS dechirper, comparing the numerical inverse Fourier transform of the impedance equation, Eq.~\ref{tildeZyq2_eq} (blue curve), with the approximation, Eq.~\ref{wq_yoff_eq} (the dashes). 
}
\label{Wq_yoff_fi}
\end{figure}

\begin{figure}[htb]
\centering
\includegraphics[width=0.6\textwidth, trim=0mm 0mm 0mm 0mm, clip]{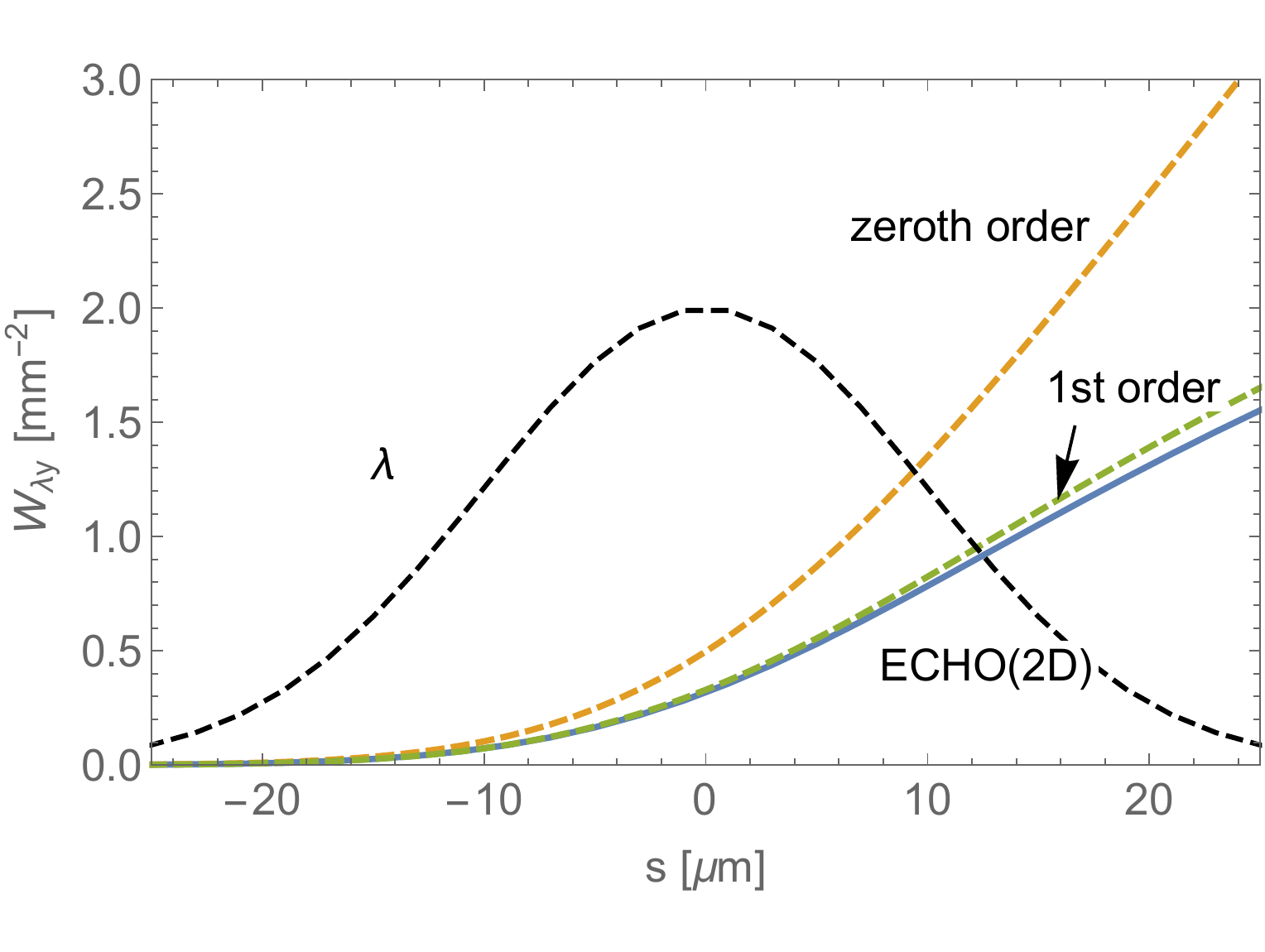}
\caption{Dipole bunch wake for a Gaussian beam with $\sigma_z=10$~$\mu$m offset in the RadiaBeam/LCLS dechirper. The offset is $y=0.5$~mm and the half-aperture is the nominal $a=0.7$~mm. Given are the numerical results of ECHO(2D) (blue), and the analytical zeroth order (red) and 1st order results (green). The bunch shape $\lambda(s)$ with the head to the left is shown in black.}
\label{Wd_gauss_yoff_fi}
\end{figure}

\begin{figure}[htb]
\centering
\includegraphics[width=0.6\textwidth, trim=0mm 0mm 0mm 0mm, clip]{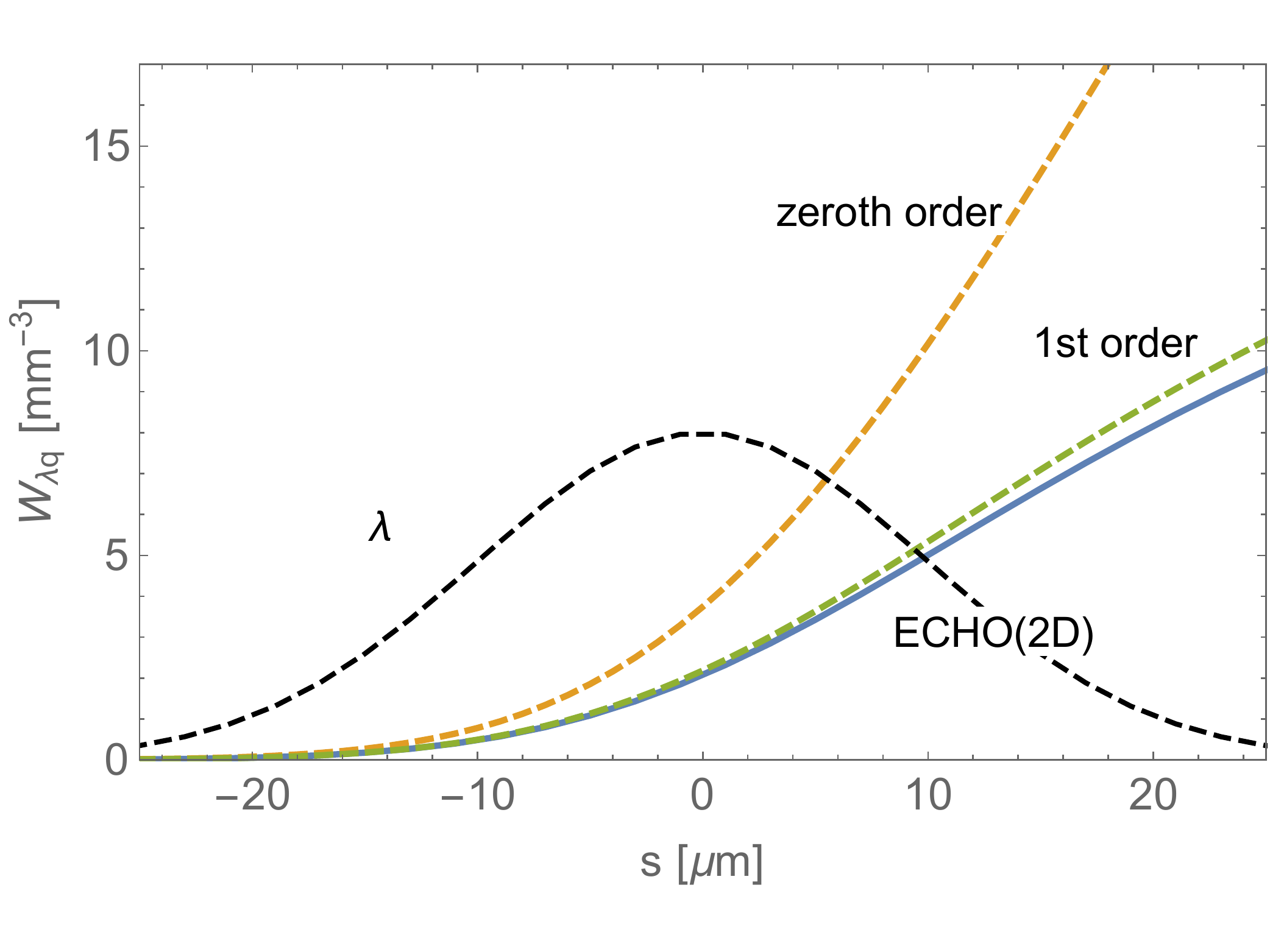}
\caption{Vertical quad bunch wake for a Gaussian beam with $\sigma_z=10$~$\mu$m offset in the RadiaBeam/LCLS dechirper. The offset is $y=0.5$~mm and the half-aperture is the nominal $a=0.7$~mm. Given are the numerical results of ECHO(2D) (blue), and the analytical zeroth order (red) and 1st order results (green). The bunch shape $\lambda(s)$ with the head to the left is shown in black.}
\label{Wq_gauss_yoff_fi}
\end{figure}

For the RadiaBeam/LCLS dechirper nominal parameters, for the quad and dipole wakes for a beam offset by $y=0.5$~mm in the vertical unit, we compared the numerical inverse Fourier transform of the impedance equations, Eq.~\ref{tildeZy2_eq} with $x_0=x$ and $y_0=y$ for the dipole case, and Eq.~\ref{tildeZyq2_eq} for the quad case, with the approximations, Eqs.~\ref{wqd_yoff_eq} and \ref{wq_yoff_eq}. The results are shown in Fig.~\ref{Wd_yoff_fi} and \ref{Wq_yoff_fi}, with the numerical results given in blue, and the analytical ones by red dashes. In both cases we start to see significant deviation for $s\gtrsim 0.15s_{0r}$. Finally, in Figs.~\ref{Wd_gauss_yoff_fi}, \ref{Wq_gauss_yoff_fi}, we compare the analytical dipole and quad bunch wakes for a $\sigma_z=10$~$\mu$m Gaussian beam with ECHO(2D) numerical calculations. We see that the agreement with the analytical model (``first order") is very good.

\subsection*{LCLS Example Calculations}

In Table~II a typical combination of LCLS beam and machine parameters is given. The beam is approximately uniform, with charge $Q=150$~pC, current $I=1.5$~kA, and full bunch length $\ell=30$~$\mu$m. The bunch wake for a uniform distribution is related to the point charge wake by
\begin{equation}
W_{\lambda }(s)=\frac{1}{\ell}\int_0^sds'\,w(s')\ ,\label{Wlambda_eq}
\end{equation}
with the beam head at $s=0$.
The voltage induced in the bunch is (in MKS units) 
\begin{equation}
V(s)=-\left(\frac{Z_0c}{4\pi}\right){QL}W_{\lambda l}(s)\ .
\end{equation}
For the zeroth order calculation, with the beam on axis, the voltage induced in the bunch tail $V(\ell)=-\pi Z_0cQL/(16a^2)$; in a 2-m section of dechirper at nominal setting (see Table~I) this equals $-13.6$~MV; for both dechirper sections combined, the induced voltage will be $-27.2$~MV. However, because of the droop of the wake, the induced chirp will not be linear, and the relative induced chirp within the bunch is reduced by the factor (see Eq.~\ref{Wl_eq})
\begin{equation}
\delta V(s)=2\frac{s_{0l}}{s}\left[1-e^{-\sqrt{s/s_{0l}}}\left(1+\sqrt{\frac{s}{s_{0l}}}\right)\right]\ .
\end{equation}
Substituting fom Eqs.~\ref{s0r_eq}, \ref{s0l_eq}, we find that $s_{0l}=430$~$\mu$m, and at the tail of the bunch $\delta V(\ell)=0.84$. The induced voltage in the tail (for the two dechiper sections combined) is $-22.8$~MV.

\begin{table}[hbt]
   \centering
   \caption{Selected LCLS beam and machine properties (at the dechirper) used in example calculations. This is a typical combination of parameters found in Ref.~\cite{Zhang15}. }
   \begin{tabular}{||l|c|c||}\hline 
        {Parameter name} & {Value}  &  Unit\\ \hline\hline 
      Beam energy, $E$       &6.6  &GeV \\
       Charge per bunch, $Q$       &150  &pC  \\
      Beam current, $I$       &1.5  &kA  \\
       Full bunch length, $\ell$       &30  &$\mu$m \\
      Normalized emittance, $\epsilon_{xn}$ / $\epsilon_{yn}$      &0.77 / 0.39  &$\mu$m  \\
      Beta function, $\beta_{x}$ / $\beta_{y}$      &4.5 / 23.7  &m  \\
      Beam size, $\sigma_{x}$ / $\sigma_{y}$      &16 / 27  &$\mu$m  \\
         \hline \hline 
   \end{tabular}
   \label{table1_tab}
\end{table}

For the on-axis quad wake, note that the bunch quad wake gives the inverse focal length, 
\begin{equation}
f_{q}^{-1}(s)=\mp\left(\frac{Z_0c}{4\pi}\right)\frac{eQLW_{\lambda q}(s)}{E}\ ,\label{finv_eq}
\end{equation}
which represents focusing in $x$ and defocusing in $y$ (assuming the corrugated plates are aligned horizontally).
The quantity $\beta_{x,y}f_{q}^{-1}(s)$, with $\beta_{x,y}$ the lattice beta function, is a measure of the lattice mismatch; if (in absolute value) it is large compared to 1, then the beam slice at $s$ will tend to be significantly mismatched. For the zeroth order wake approximation [let $s_{0q}\rightarrow\infty$ in Eq.~\ref{wqd2_eq} and insert into Eq.~\ref{Wlambda_eq}]
\begin{equation}
W_{\lambda q}(s)=\left(\frac{\pi^4}{64a^4}\right)\frac{s^2}{\ell}\ .
\end{equation} 
According to the zeroth order calculation, at the tail of the bunch, $\beta_xf_q^{-1}(\ell)=-0.35$ and $\beta_yf_q^{-1}(\ell)=1.84$. The relative droop in the bunch wake, which we define as the bunch wake divided by its zeroth order approximation, is (see Eqs.~\ref{wqd2_eq}, \ref{Wlambda_eq})
\begin{equation}\label{ydroop_eq}
\delta V_q(s)=4(s_{0q}/s)^2\left[2e^{-\sqrt{s/s_{0q}}}\left(3\left[1+\sqrt{s/s_{0q}}\right]+s/s_{0q}\right)+s/s_{0q}-6\right]\ .
\end{equation} 
Here $s_{0q}=170$~$\mu$m, and the wake droop is $\delta V_q(\ell)=0.80$.
Thus $\beta_xf_q^{-1}(\ell)=-0.28$ and $\beta_yf_q^{-1}(\ell)=1.47$; the mismatch in $y$ at the bunch tail is significant.

There is interest in passing the beam close to one dechirper jaw, to induce a strong kick variation along the beam. With the beam offset from the $y$-axis in the vertical unit, the transverse voltage at position $s$ within a uniform bunch, in the zeroth order approximation, is (in [V], see Eq.~\ref{wqd_yoff_eq})
\begin{equation}
V_{yd}(s)=\frac{\pi^2}{64} {Z_0c}\frac{QLs^2}{a^3\ell}\sec^2(\frac{\pi y}{2a})\tan(\frac{\pi y}{2a})\ .
\end{equation}
Note that the kick varies quadratically, not linearly, with $s$.
For our example parameters, with the beam offset by $y=0.5$~mm, we find that, in the tail of the bunch and in the zeroth order approximation, the kick is $V_{yd}(\ell)=5.0$~MV. The relative droop in the wake of the uniform bunch is given by Eq.~\ref{ydroop_eq}, but using $s_{0yd}$ (Eq.~\ref{s0yd_eq}) as the scale factor. Here $s_{0yd}=28$~$\mu$m and the relative droop $\delta V_{yd}=0.59$. Thus the kick at the tail of the bunch is  $V_{yd}(\ell)=3.0$~MV. 
Note that to achieve a 3~MV voltage differential over a length of 30~$\mu$m with an X-band transverse cavity (of frequency 11.4~GHz), would require a peak voltage of 420~MV.

As the beam moves off axis, the dipole kick increases but so will the quad wake; in fact, it increases more rapidly.
In this case, the quad bunch wake, according to the zeroth order approximation, is given by (see Eq.~\ref{wq_yoff_eq})
\begin{equation}
W_{\lambda q}(s)=\frac{\pi^4}{64} \frac{s^2}{a^4\ell}\left[2-\cos(\frac{\pi y}{a} ) \right]\sec^4(\frac{\pi y}{2a}) \ .
\end{equation}
Consider again the beam offset by $y=0.5$~mm. Combining the above equation with Eq.~\ref{finv_eq}, we find that, at the tail of the bunch in the zeroth order approximation, $\beta_xf_q^{-1}(\ell)=-26$ and $\beta_yf_q^{-1}(\ell)=136$. The relative droop in the bunch wake is given by Eq.~\ref{ydroop_eq}, but using as scale factor $s_{0yq}$ (Eq.~\ref{s0yq_eq}).
Here $s_{0yq}=16$~$\mu$m, and the wake droop is $\delta V_q=0.50$.
Thus $\beta_xf_q^{-1}(\ell)=-13$ and $\beta_yf_q^{-1}(\ell)=68$. These numbers still represent large optics mismatch. One can obtain partial compensation by also passing the beam through the horizontal dechirper unit---with half gap set to $a=0.7$~mm---at offset $x=0.5$~mm, though, with such large mismatch, substantial compensation will likely be difficult to achieve.

\subsection*{Discussion/Conclusions}

Flat corrugated structures, to be used as ``dechirpers" in linac-based X-ray FELs, have been built and tested at several laboratories around the world. Recently, interest has grown in using these devices also as fast kickers, by passing the beam close to one jaw of a dechirper section, in order to excite strong transverse wakefields to kick the beam tail. 
In this report we develop analytical models of the longitudinal and transverse wakes, on and off axis for realistic structures, and then compare them with two numerical calculations: (i)~numerical integration of the general impedance formulas that assumed a surface impedance, and (ii)~time domain, finite difference calculation using ECHO(2D). We generally find good agreement. These analytical models (that we call ``first order" formulas) approximate the droop at the origin of the longitudinal wake and of the slope of the transverse wakes; they represent an improvement in accuracy over earlier, ``zeroth order" formulas given in \cite{Bane16}. 

The formulas developed here can be useful for parameter studies and beam dynamics simulations. They seem to be quite accurate near the dechirper axis, for distances $s\lesssim0.25s_{0r}$, where $s_{0r}$ is a scale factor for the structure. But as the beam moves off axis, the range over which the formulas are accurate becomes shorter. 


In this report we, in addition, performed example calculations for the RadiaBeam/LCLS dechirper that has been recently installed in LCLS. The bunch distribution was taken to be uniform with full length $\ell=30$~$\mu$m, charge $Q=150$~pC, and energy $E=6.6$~GeV. With dechirper half-gap $a=0.7$~mm and the beam on axis, the energy loss at the bunch tail (for the two dechirper sections combined) is reduced from 27~MeV to 23~MeV (a drop of 16\%) by the wake droop. With the beam moved to $y=0.5$~mm in the vertical dechirper unit, the kick at the bunch head is zero and at the bunch tail is 3~MV, which is a large differential over a 30~$\mu$m length. The mismatch caused by the quad wake, however, is also large: at the bunch tail $\beta_xf_q^{-1}(\ell)=-13$ and $\beta_yf_q^{-1}(\ell)=68$. One can obtain partial compensation by also passing the beam through the horizontal unit at offset $x=0.5$~mm, though substantial compensation will likely be difficult to achieve.

\subsection*{Acknowledgements}

We thank the team commissioning the RadiaBeam/LCLS dechirper, led by R. Iverson, for showing us how the dechirper affects the beam in practice, and J. Zemella, who has worked with us and performed numerical calculations of the dechirper wakes. Work supported by the U.S.
Department of Energy, Office of Science, Office of Basic
Energy Sciences, under Contract No. DE-AC02-76SF00515.

\end{document}